\documentclass[twocolumn,aps,superscriptaddress,pra,showpacs]{revtex4}
\usepackage{epsfig} 
\newcommand{\ket}[1]{|#1\rangle} \newcommand{\bra}[1]{\langle#1|}

\newcommand{\eq}{\begin{equation}} \newcommand{\fine}{\end{equation}}

\begin{document}

\title{Complete analysis of measurement-induced entanglement localization on a three-photon system}

\author{Miroslav Gavenda}
\author{Radim Filip}
\affiliation{Department of Optics,
Palack\' y University, 17.~listopadu 12, 771~46 Olomouc, Czech Republic}

\author{Eleonora Nagali}
\affiliation{Dipartimento di Fisica, Sapienza Universit\`a di Roma, piazzale
Aldo Moro 5, 00185 Roma, Italy}

\author{Fabio Sciarrino} 
\affiliation{Dipartimento di Fisica, Sapienza Universit\`a di Roma, piazzale
Aldo Moro 5, 00185 Roma, Italy}
\affiliation{Istituto Nazionale di Ottica Applicata, Largo E. Fermi 6,
50125 Firenze, Italy}

\author{Francesco De Martini} 
\affiliation{Dipartimento di Fisica, Sapienza Universit\`a di Roma, piazzale
Aldo Moro 5, 00185 Roma, Italy}
\affiliation{Academia Nazionale dei Lincei, Via della Lungara 10, 00165
Roma, Italy}


\begin{abstract}
We discuss both theoretically and experimentally elementary two-photon
polarization entanglement localization after break of entanglement caused
by linear coupling of environmental photon with one of the system photons. The
localization of entanglement is based on simple polarization measurement of the
surrounding photon after the coupling. We demonstrate that non-zero
entanglement can be localized back irrespectively to the distinguishability of coupled photons. 
Further, it can be increased by local
single-copy polarization filters up to an amount violating Bell
inequalities. The present technique allows to restore entanglement in that cases, when the entanglement distillation does not produce any entanglement out of the coupling.
\end{abstract}


\maketitle

\section{Introduction}
Quantum entanglement, the fundamental resource in quantum information
science, is extremely sensitive to coupling to surrounding
systems. By this coupling the entanglement is reduced or even completely
vanishes. In the
case of partial entanglement reduction, quantum
purification and distillation protocols can be adopted \cite{Benn96,Deut96,Zhao03,Pan03,Walt05}. In
quantum distillation protocols, entanglement can be increased
without any operation on the surrounding systems. Using a collective procedure, the distillation
probabilistically transforms many copies of less entangled states to a
single more entangled state, using a collective procedure just at the
output of the coupling \cite{Kwia01}. Fundamentally, the distillation requires at least a bit of residual
entanglement passing through the coupling. 

When the coupling with the environment completely destroys the entanglement, 
both purification and distillation protocols can not be exploited. 
Hence another branch of methods like unlocking of hidden 
entanglement \cite{Cohe98} or entanglement localization \cite{DiVi04,Vers04,Smol05,Popp06,Gour06}
must be used.  
In reference \cite{Cohe98} there has been estimated that in order to unlock the hidden entanglement from a
separable mixed state of two subsystems an additional bit or bits of
classical information is needed to determine which entangled state is
actually present in the mixture.
However no strategy was presented in order to extract this information
from the system of interest or the environment. 
Latter, to retrieve the entanglement, the localization protocols \cite{DiVi04,Vers04,Smol05,Popp06,Gour06} 
have been introduced working on the principle of getting some nonzero amount of bipartite entanglement
from some multipartite entanglement. That means  performing some kind of
operation on the multipartite system (or on its part) may induce bipartite
entanglement between subsystems of interest.

We can say that the entanglement breaks because it 
is so inconveniently redistributed among system of interest 
and environment that it is transformed into generally complex multipartite entanglement. 
Entanglement can be localized back for the
further application just by performing suitable measurement on the surrounding system
and the proper feed-forward quantum correction. The measurement and
feed-forward operation substitute, at least partially, a full inversion
of the coupling which requires to keep very precise interference with
the surrounding systems. 
For many-particle systems, the maximal value of
the localizable entanglement depends on the coupling and also on initial
states of surrounding systems. 

\begin{figure}
\centerline{\psfig{width=7.0cm,angle=0,file=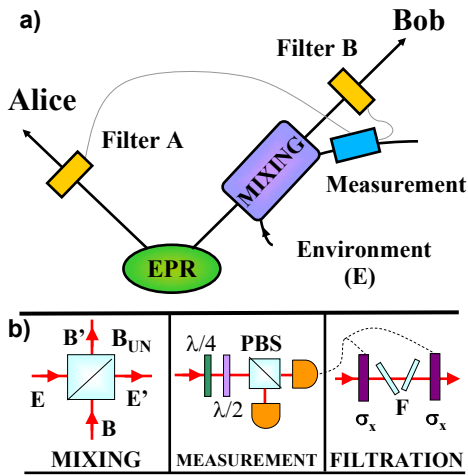}}
\caption{(Color online) \textbf{a)} Schematic representation of the localization protocol. \textbf{b)} Detailed representation of the three steps of the localization procedure.}
\label{coherentfig1}
\end{figure}

To understand the mechanism of the entanglement localization, we simplify the complex many
particle coupling process to a set of sequential couplings with a
single particle representing an elementary surrounding system $E$
\cite{Zure03}: see FIG.~\ref{coherentfig1}. Then, our focus is just on the
three-qubit entanglement produced by this elementary coupling of one qubit from
maximally entangled state of two qubits $A$ and $B$ to a third surrounding
qubit $E$. Before the coupling to the qubit $B$, we have typically no control on the quantum state of qubit $E$, hence considered to be unknown. Further, the qubit $E$ is typically not interfering or
weakly interfering with the qubit $B$.   
Recently Sciarrino et al. \cite{Sciar09} reported the entanglement localization after the coupling to an incoherent noisy system.\\
In this paper, we report the study in depth of the key role of two-photon {\em
interference} in the process of entanglement localization presented in
our previous paper \cite{Sciar09}. In particular we focus on the
theory of the localization protocol, both considering a distinguishable  and
an indistinguishable photon coupling with the environment. The experimental
implementation of such processes will be presented as well, with large
attention to the experimental setup. In order to carry out our
analysis on the dynamic of the localization protocol, we have
considered a simple process for the interaction between the signal
photon and the surrounding one. To be specific, we assume a
polarization insensitive linear coupling between distinguishable or
partially distinguishable surrounding
unpolarized single photon $E$ and
single photon $B$ from the photon pair $A,B$ maximally entangled in
the polarization degree of freedom. After this coupling, the state
of
two entangled photons $A,B$ turns to be a mixture of the two-photon maximally
entangled state and the separable two unpolarized photons. For some value of
the coupling, this state gets separable and the entanglement is completely
redirected to the surrounding system. We theoretically prove that for any
linear coupling, it is always possible to localize a non-zero entanglement back to
the photon pair $A,B$ just by a polarization sensitive detection of
the photon $E$. Interestingly, this qualitative result can be obtained
irrespectively to the level of distinguishability and noise in
the surrounding photon $E$. We discuss the impact of distinguishability in the
entanglement localization more in details and give physical explanation
of the differences between the localization for distinguishable or partial
distinguishable photon $E$.
The performance of this localization can be further enhanced using a simple
single-copy distillation method \cite{Horo97,Lind98,Kent99,Horo99,Vers01,Cen02,Vers02,Pete04,Wang06,Lian08}, up to a state violating
Bell inequalities \cite{Bell64,Clau69}, independently on the noise and
distinguishability
of the photon $E$.

The paper is organized as follows. In Sec.~II we discuss theory of the polarization entanglement localization, for a comparison, first for
the coupling of fully indistinguishable photons (Sec.~ II.A) and then for
realistic linear coupling of partially distinguishable photons (Sec.~II.B). In Sec.~III, the 
experimental results are discussed in details. 
The Sec.~III.A describes the experimental setup and the state preparation. 
The Sec.~III.B contains the experimental result about the localization for
the coupling of fully distinguishable photons, the Sec.~III.C presents the
results of localization for the coupling of indistinguishable photon whereas the
Sec.~III.D reports on results of localization for the coupling of partially
indistinguishable photons. The summary is done in the Conclusion. In the Appendix A, a modification of the entanglement localization for the polarizing linear coupling is described.

\section{Theory}
Let us assume that we generate a maximally entangled polarization state
$|\Psi_-\rangle_{AB}=(|HV\rangle-|VH\rangle)/\sqrt{2}$ of two photons $A$
and $B$. The form of the maximally entangled state is not relevant, as the same results can be obtained for
any maximally entangled state. The first photon is kept by Alice and second is coupled
to the surrounding photon $E$ in an unknown state, $\rho_E=\openone/2$
(completely unpolarized state), by a simple linear polarization
insensitive coupling (a beam splitter with transmissivity $T$). After
the coupling, three possible situations can be observed: both photons
$B$ and $E$ go simultaneously to either mode $k'_B$ or mode $k'_E$,
or only a single photon is separately presented in both output modes
$k'_B$ or $k'_E$ (see FIG.~\ref{coherentfig1}). We will focus only on the last case, that is a single
photon output from the coupling. In this case, it is in principle not
possible to distinguish whether the output photon is the one from the entangled
pair ($B$) or the unpolarized photon $E$. We investigate in
detail, the impact of the distinguishability of the surrounding system on the
localization procedure. Note that the system $E$ can be also produced by
the entanglement source and then coupled to the entangled state
through a subsequent propagation. Therefore, it is not fundamentally
important whether $E$ is produced by truly independent source or not.

\subsection{Coupling of indistinguishable photons}

To understand the role of partial distinguishability in the entanglement
localization, let us first assume that both photons $B$ and $E$ are perfectly indistinguishable. Then, if both the photons
leave the linear polarization-insensitive coupling separately, the coupling transformation can
be written for any polarization state $|\Psi\rangle_B$ as:
\begin{eqnarray}\label{coupling}
|\Psi\rangle_B|\Psi\rangle_E
&\rightarrow & (T-R)|\Psi\rangle_B|\Psi\rangle_E,\nonumber\\
|\Psi\rangle_B|\Psi_{\bot}\rangle_E &\rightarrow &
T|\Psi\rangle_B|\Psi_{\bot}\rangle_E-R|\Psi_{\bot}\rangle_B|\Psi\rangle_E.
\end{eqnarray}
Both the singlet state $|\Psi_-\rangle_{AB}$
as well as $\rho_E$ can be written in the general
basis $|\Psi\rangle$ and $|\Psi_{\bot}\rangle$, namely:
$|\Psi_-\rangle_{AB}=(|\Psi\Psi_{\bot}\rangle_{AB}-
|\Psi_{\bot}\Psi\rangle_{AB})/\sqrt{2}$
and
$\rho_E=(|\Psi\rangle_E\langle\Psi|+|\Psi_{\bot}\rangle_E\langle\Psi_{\bot}|)/2$.
Therefore, it is simple to find that the output state $\sigma_{ind}^{I}$ (tracing
over photon $E$) is exactly a mixture of the entangled state with the
unpolarized noise (Werner state) \cite{Wern89}:
\begin{equation}\label{W}
  \sigma_{ind}^{I}=\frac{4F_{ind}-1}{3}|\Psi_-\rangle_{AB}\langle\Psi_-|+\frac{(1-F_{ind})}{3}\openone_{A}\otimes
  \openone_{B},
\end{equation}
where the fidelity $F_{ind}$, that is the overlap with
the singlet state $|\Psi_-\rangle_{AB}$, takes values $0 \leq F_{ind} \leq 1$ and reads:
\begin{equation}
F_{ind}=\frac{(1-3T)^2}{4\left(1-3T(1-T)\right)}.
\end{equation}
The state is entangled only if $1/2<F\leq 1$ and the concurrence
\cite{Woot98} reads $C_{ind}=2F_{ind}-1$. 
If
$F_{ind}<1/2$ the entanglement is completely lost and the state (\ref{W})
is separable. The explicit formula for concurrence is
\begin{equation}
C^{I}_{ind}=\mbox{max}\left(0,\frac{3T^2-1}{2\left(1-3T(1-T)\right)}\right).
\end{equation}
The probability of this situation is
$P^{I}_{ind}=\left(T^2+(T-R)^2+R^2\right)/2$ and the condition for
entanglement breaking channel is $T<1/\sqrt{3}$. In order to achieve a maximal violation of Bell inequalities \cite{Horo95}, we have to refer to the quantity $B^{I}_{ind}$ given by the expression
\begin{equation}\label{bmax1}
  B^{I}_{ind}=\frac{2\sqrt{2}T|1-2T|}{1-3T(1-T)}
\end{equation} 
The violation appears only for $T>0.68$, imposing a condition on the coupling even more strict.

More generally, all the result that will be presented in the paper
can be directly extended for general passive coupling between two
modes. It can be represented by Mach-Zehnder interferometer consisting
to two unbalanced beam splitters (with transmissivity $T_1$ and
$T_2$) and two phase shifts $-\phi$ and $\phi$ separately in arms
inside the interferometer. All the results depends only on a single
parameter: the probability $T$ that the entangled photon will leave the
coupling as the signal, which is explicitly equal to:
\begin{equation}
T=T_1+T_2-2T_1T_2+\sqrt{T_1T_2R_1R_2}\cos 2\phi.
\end{equation} Thus
all results, here simply discussed for a beam splitter, are valid for
any passive coupling between two photons.

After the measurement of polarization on photon $E$, let us say by the
projection on an arbitrary state $|\Psi\rangle_E$, the state of photons
$A,B$ is transformed into
\begin{equation}\label{rec}
\sigma^{II}_{ind}=\frac{1}{P^{II}_{ind}}\left((T^2+(T-R)^2)|T\rangle\langle
T|+R^2|\Psi_{\bot}\Psi_{\bot}\rangle\langle\Psi_{\bot}\Psi_{\bot}|
\right)
\end{equation}
where the state $|T\rangle$ represents the unbalanced singlet state:
\begin{equation}
|T\rangle=\frac{1}{\sqrt{T^2+(T-R)^2}}\left(T|\Psi\Psi_{\bot}\rangle
-(T-R)|\Psi_{\bot}\Psi\rangle \right)
\end{equation}
and $P^{II}_{ind}=T^2+(T-R)^2+R^2$. Calculating the concurrence
as measure of entanglement, we found
\begin{equation}
C^{II}_{ind}=\frac{T|2T-1|}{1-3(1-T)T} 
\end{equation}
which is always larger than zero if $T\not=0,1/2$.
It is worth notice that, irrespective to polarization noise in the qubit $E$, it is possible to probabilistically
localized back to the original photon pair $A$ and $B$ a non-zero entanglement for all $T\not=0,1/2$.

The basic principle of this process can be understood by comparing Eq.~(\ref{W})
and Eq.~(\ref{rec}). The unpolarized noise in Eq.~(\ref{W}) has been
transformed to a fully correlated polarized noise in Eq.~(\ref{rec}) just
by the measurement on $E$. This measurement can be arbitrary and does
not depend on the coupling strength $T$. Thus, it is not necessary to
estimate the channel coupling before the measurement. The positive result comes from the observation
that a fully correlated and polarized noise is less destructive to maximal entanglement
than the completely depolarized noise. Further, the entanglement
localization effect persists even if an arbitrary phase or amplitude
damping channel affects photon $E$ after the coupling. If there is still
a preferred basis in which the same classical correlation can be kept,
it is sufficient for the entanglement localization. Unfortunately,
the localized state is entangled but the localization itself is not
enough to produce a state violating Bell inequalities. The maximal Bell
factor $B_{ind}^{II}$ is identical to (\ref{bmax1}) although the state
has completely changed its structure.

However if $T$ is known then the entanglement of state (\ref{rec}) and the Bell
inequality violation can be further increased by the single-copy
distillation \cite{Horo97,Lind98,Kent99}. First, a local polarization filter
can be used to get the balanced singlet state in the mixture. An
explicit construction of the single local filter at Bob's side
is given by $|\Psi\rangle\rightarrow (T-R)/T
|\Psi\rangle,|\Psi_{\bot}\rangle\rightarrow |\Psi_{\bot}\rangle$ for
$T>1/3$. If $T<1/3$ then a different filtering $|\Psi\rangle\rightarrow
|\Psi\rangle,|\Psi_{\bot}\rangle\rightarrow T/(T-R) |\Psi_{\bot}\rangle$
has to be applied. Thus the Werner state (\ref{W}) is conditionally
transformed to a maximally entangled mixed state (MEMS)\cite{Vers01b,Ishi00,Wei03,Cine04}:
\begin{equation}
\label{state1a} 
\frac{\left((T-R)^2|\Psi_-\rangle\langle
\Psi_-|+R^2|\Psi_{\bot}\Psi_{\bot}\rangle\langle\Psi_{\bot}\Psi_{\bot}|
\right)}{(T-R)^2+R^2}
\end{equation}
where the probability of success is given by $(T-R)^2+R^2$. Due to symmetry, the same situation arises
if the state $|\Psi_{\bot}\rangle$ is detected, just replacing the state
$\Psi\leftrightarrow \Psi_{\bot}$. If the local filtering on Bob's side is
applied according to the projection, it is possible to reach the MEMS with
twice of success probability. Now, a difference between Eq.(\ref{state1a})
and Eq.~(\ref{W}) is only in an additive fully correlated polarized
noise. Such result can be achieved by an operation on the photon $B$, while
a processing on other photon is not required.


Further local filtration performed on both photons $A$ and $B$
can give maximal entanglement achievable from a single copy. Both
Alice and Bob should implement the filtration $|\Psi\rangle\rightarrow
|\Psi\rangle$ and $|\Psi_{\bot}\rangle\rightarrow
\sqrt{\epsilon}|\Psi_{\bot}\rangle$, where $0<\epsilon\leq 1$.  The filtered state is then
\begin{eqnarray}
  \sigma^{III}_{ind}&=&\frac{1}{4P^{III}_{ind}}\left(\epsilon
(2T-1)^2|\Psi_-\rangle\langle\Psi_-|+\right.\nonumber\\
& &\left.\epsilon^2(1-T)^2
|\Psi_{\bot}\Psi_{\bot}\rangle\langle\Psi_{\bot}\Psi_{\bot}|\right)
\end{eqnarray} where
$P^{III}_{ind}=\left(\epsilon(2T-1)^2+\epsilon^2(1-T)^2\right)/4$
is probability of success. The state has the
following concurrence for $T\not=0,1/2$
\begin{equation}
C^{III}_{ind}(\epsilon)=\frac{1}{1+\epsilon\frac{(1-T)^2}{2(1-2T)^2}}
\end{equation}
As $\epsilon$ goes to zero, the concurrence approaches unity, except
for $T=0,1/2$, where it vanishes. Thus we can get entanglement arbitrarily
close to its maximal value for any $T$ (except $T=0,1/2$). For $T=1/2$,
it could not appear due to bunching effect between the photons at the
beam splitter. Such result can be achieved since we can completely
eliminate the fully correlated polarized noise, which is actually
completely orthogonal to the single state $|\Psi_-\rangle$. We stress once again that
maximal entanglement was probabilistically approached irrespective to the
initial polarization noise of the photon $E$, just by its single-copy
localization measurement and single-copy distillation.

If $\epsilon \rightarrow 0$ we get simply $B_{ind}^{III}=2\sqrt{2}$
for any $T$, except $T=0,1/2$. Since the maximal entanglement allows ideal teleportation
of unknown state we can conclude that the linear coupling to the
distinguishable
photon even in an unknown state could be practically probabilistically
reversed. It is interesting to make a comparison with the
deterministic localization procedure. To achieve it, we need perfect
control of the state of photon $E$ before the coupling. We can
know it either a priori or by a measurement on the more
complex systems predicting pure state of $E$. Being $|\Psi\rangle_E$
the state of $E$ after coupling, three-photon state is proportional to
$|\Psi\rangle_A\left(T|\Psi_{\bot}\Psi\rangle_{BE}-R|\Psi\Psi_{\bot}\rangle_{BE}\right)-(T-R)
|\Psi_{\bot}\Psi\Psi\rangle_{ABE}$. This belongs to not-symmetrical W-state,
rather than to GHZ states, therefore the maximal entanglement can
not be obtained deterministically. It can be done by the projection on
$|\Psi\rangle_E$ followed by the single-copy distillation. From this point
of view, the unpolarized noise of photon $E$ does not qualitatively change
the result of the localization, but only decreases the success rate.

\begin{figure}
\centerline{\psfig{width=8.0cm,angle=0,file=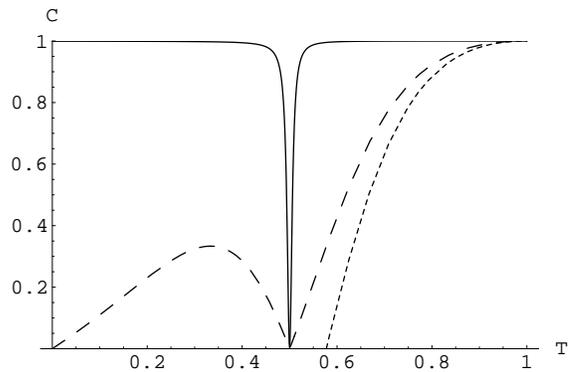}}
\caption{Concurrence for indistinguishable
photon $E$ ($p=1$): $C^{I}_{ind}$ without measurement (dotted),
$C^{II}_{ind}$ with measurement (dashed), $C^{III}_{ind}$
with measurement and LOCC filtration (full), for $\epsilon=0.005$.}
\label{coherentfig}
\end{figure}

\begin{figure} \centerline{\psfig{width=8.0cm,angle=0,file=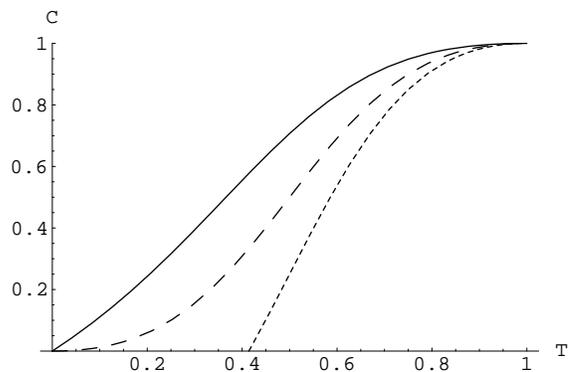}}
\caption{Concurrence for distinguishable photon ($p=0$): $C^{I}_{dis}$
without measurement (dotted), $C^{II}_{dis}$ with measurement (dashed),
$C^{III}_{dis}$ with measurement and LOCC filtration (full)}
\label{incoherentfig}
\end{figure}

\subsection{Coupling of partially distinguishable photons}

Up to now, we have assumed that both photons $B$ and $E$ are
indistinguishable in the linear coupling. On the other hand, if they
are partially or fully distinguishable), then it is in principle possible to perfectly filter out surrounding photons from
the signal photons after the coupling. 
Although they are in principle partially or fully
distinguishable, any realistic attempt to discriminate them would be
typically too noisy to offer such information. Here we are interested in
the opposite situation, when distinguishable photon $B$ and surrounding
photon $E$ cannot be practically distinguished after the coupling. It
is an open question whether the entanglement localization method can
still help us.
For simplicity, let us assume a mixture of the previous
case with a situation where both the photons $B$ and $E$
are completely distinguishable. The state is mixture $p\sigma_{ind}+(1-p)\rho_{dis}$, where only with probability $p$ the photons are indistinguishable. After the coupling and
tracing over photon $E$, the output state is once more
a Werner state (\ref{W}) with fidelity
\begin{equation}
F(p)=\frac{(4p+5)T^2-2(2p+1)T+1}{4\left(1-(2+p)T(1-T)\right)}
\end{equation}
and success probability $P(p)=1-(2+p)T(1-T)$. The
concurrence is given by $C(p)=2F(p)-1$.  The entanglement between
$A$ and $B$ is lost if
\begin{equation}
 p>\frac{T^2+2T-1}{2T(1-T)}
\end{equation}
and thus for $T<\sqrt{2}-1$, the entanglement between $A$
and $B$ is lost for any $p$. The maximal Bell factor
\begin{equation}
B(p)=2 \sqrt{2} T \frac{|T-p (1-T)|}{1-(p+2) (1-T) T}
\end{equation}
gives violation of Bell inequalities only if $p\in[0,1]$ satisfies
\begin{equation}
p<\sqrt{2}+\frac{1}{1-T}-\frac{1+\sqrt{2}}{T}.
\end{equation}

Let us discuss first the entanglement localization for the
fully distinguishable photon case ($p=0$). In this limit,
after the coupling the fidelity reads: 
\begin{equation}
F_{dis}=\frac{5T^2-2T+1}{4\left(1-2T(1-T)\right)},
\end{equation} 
and the concurrence and the maximal
violation of Bell inequalities are 
\begin{equation}
C^{I}_{dis}=\frac{T^2+2T-1}{2(1-2T(1-T))},\,\,\,
B^{I}_{dis}=2\sqrt{2}\frac{T^2}{T^2+(1-T)^2}.
\end{equation} 
The concurrence vanishes for $T<\sqrt{2}-1$ and the
violation of Bell inequalities disappears for $T<1+\frac{1}{\sqrt{2}}\left(1-\sqrt{1+\sqrt{2}}\right)\approx 0.608$.
In both cases, the thresholds are lower than for the coupling of
indistinguishable photons;
the distinguishability of photon $E$ leads to less robust transmission of the
quantum resources.

If the projection on $|\Psi\rangle_E$ is
performed on the environmental photon after the
coupling, the resulting state reads 
\begin{equation}
\label{incoh1}
\rho^{II}_{dis}=\frac{1}{P^{II}_{dis}}
(\frac{T^2}{2}|\Psi_-\rangle_{AB}\langle\Psi_-|+\frac{R^2}{2}|\Psi_{\bot}
\rangle_A\langle\Psi_{\bot}|\otimes \openone/2_B) 
\end{equation} 
where $P^{II}_{dis}=(T^2+R^2)/2$. Clearly, the additive noisy term in the
Eq.(\ref{incoh1}) is separable and the full correlation in the partially
polarized noise (\ref{rec}) does not arise here, in comparison with
the fully indistinguishable case. Except for $T=0$, this state is exhibiting non-zero concurrence
\begin{equation} 
C^{II}_{dis}=\frac{T^2}{2T^2-2T+1} 
\end{equation}
which is a decreasing function of $T$. The state
just after the localization does not violate Bell inequalities, since the
maximal Bell factor is the same as after the mixing $B^{II}_{dis}=B^{I}_{dis}$.

Similarly to the previous discussion, projecting on $|\Psi_{\bot}\rangle_E$,
we get an identical state, only replacing $\Psi\leftrightarrow \Psi_{\bot}$. This
means that even for completely non-interfering photons $B$ and $E$, a
polarization measurement on the photon $E$ can localize entanglement
back to photons $A$ and $B$. To approach it with maximal success rate,
local unitary transformations doing $\Psi\leftrightarrow \Psi_{\bot}$
have to be applied if $|\Psi_{\bot}\rangle$ is measured. Similarly
to previous indistinguishable case, the photon $E$ is not
affected by the amplitude or phase damping channel. It is only important to
keep a preferred basis without any noise influence.

The physical mechanism which models the localization
process for coupling of distinguishable photons is different from the previous case. Here,
the three-photon mixed state after the coupling is
proportional to 
\begin{equation}
T^2|\Psi_-\rangle_{AB}\langle\Psi_-|\otimes
\openone/2_E+R^2|\Psi_-\rangle_{AE}\langle\Psi_-|\otimes \openone/2_B
\end{equation}
describing just a random swap of the photons $B$ and $E$ at the beam
splitter. Although there is no correlation generated by the
coupling between photons $A$ and $B$, contrary to previous case, still
a correlation between photons $A$ and $E$ will appear in the second
part of expression due to the initial entangled state. It is used to
conditionally transform the locally completely
depolarized state of photon $A$ to a fully polarized state. Of course,
it does not have any impact on the local state of photon $B$. But it
is fully enough to localize entanglement for any $T$ after the projection
on photon  $E$. It is rather a non-local correlation effect where the polarization
noise is corrected not at photon $B$ but at photon $A$.

The entanglement can be further enhanced using local filtration
and classical communication; in the limit $\epsilon \rightarrow 0$ and
introducing optimal filtering given
by \begin{eqnarray} |\Psi_{\bot}\Psi\rangle &\rightarrow &
\frac{T}{\sqrt{2T^2-2T+1}}|\Psi_{\bot}\Psi\rangle,\nonumber\\
|\Psi_{\bot}\Psi_{\bot}\rangle &\rightarrow &\epsilon
|\Psi_{\bot}\Psi_{\bot}\rangle \end{eqnarray} (whereas all other basis
states are preserved)  
the final state approaches  
\begin{eqnarray}\label{rho_III_dis}
\rho^{III}_{dis}&=&\frac{1}{2}|\Psi\Psi_{\bot}\rangle_{AB}\langle\Psi\Psi_{\bot}|+\frac{1}{2}|\Psi_{\bot}\Psi\rangle_{AB}\langle\Psi_{\bot}\Psi|
\nonumber \\ &-&
\frac{T}{2\sqrt{2T^2-2T+1}}\left(|\Psi\Psi_{\bot}\rangle\langle\Psi_{\bot}\Psi|+
\text{h.c.} \right).
\end{eqnarray}
The filtrations turns the state (\ref{incoh1}) into
state (\ref{rho_III_dis}) which appears as a result of the action of phase damping channel on
the input singlet state. The filtrations eliminated the noise component
$|\Psi_{\bot}\Psi_{\bot}\rangle_{AB}\langle\Psi_{\bot}\Psi_{\bot}|$ from
the state (\ref{incoh1})
and this ensures the violation of Bell inequalities for the final state (\ref{rho_III_dis}).  

The concurrence approaches $C^{III}_{dis}=\sqrt{C^{II}_{dis}}$, at cost of a
decreasing success rate. 
But, at contrary to the case of two perfectly
interfering photons, maximal entanglement cannot be approached for any
$T<1$. This is caused by lack of interference in the entanglement localization
for distinguishable photons. Of course, the collective protocols can
be still used, since all entangled two-qubit states are distillable to
maximally entangled state \cite{Horo97}.

A maximal value of the Bell factor 
\begin{equation}
B^{III}_{dis}=2\sqrt{1+\frac{T^2}{T^2+(1-T)^2}} 
\end{equation}
can be achieved if $\epsilon\rightarrow 0$ for any T. 
Although Bell inequalities are violated in this case for any $T>0$
the full entanglement cannot be recovered
at any single copy.

Also here it is instructive to make a comparison with the
case in which the photon $E$ is in the known pure state
$|\Psi\rangle_E$. Then after the coupling the three-photon state
is proportional to $T^2|\Psi_-\rangle_{AB}\langle\Psi_-|\otimes
|\Psi\rangle_E\langle\Psi|+R^2|\Psi_-\rangle_{AE}\langle\Psi_-|\otimes
|\Psi\rangle_B\langle\Psi|$ and after the projection on $|\Psi\rangle_E$,
it transforms to $T^2|\Psi_-\rangle_{AB}\langle\Psi_-|+
\frac{1}{2}R^2|\Psi_{\bot}\rangle_A\langle\Psi_{\bot}|\otimes
|\Psi\rangle_B\langle\Psi|$. Since both the contributions are not
orthogonal there is no local filtration procedure which could filter out
the $|\Psi_-\rangle_{AB}$ from the noise contribution. Therefore, although
we could have a complete knowledge about pure state of the photon $E$,
its distinguishability does not allow to achieve maximal entanglement from the
single copy distillation.

\begin{figure}
\centering
\includegraphics[scale=.4]{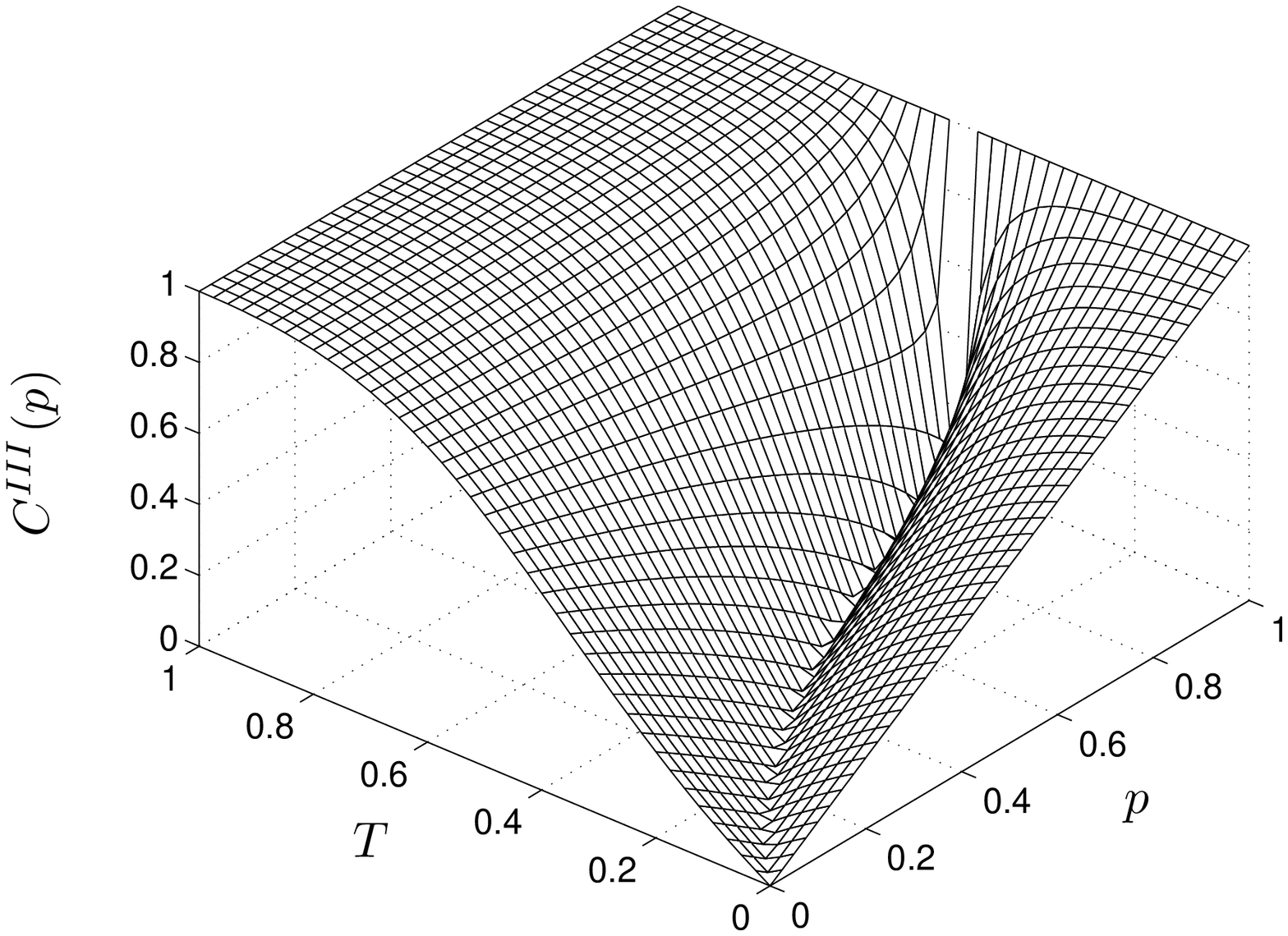}
\includegraphics[scale=.4]{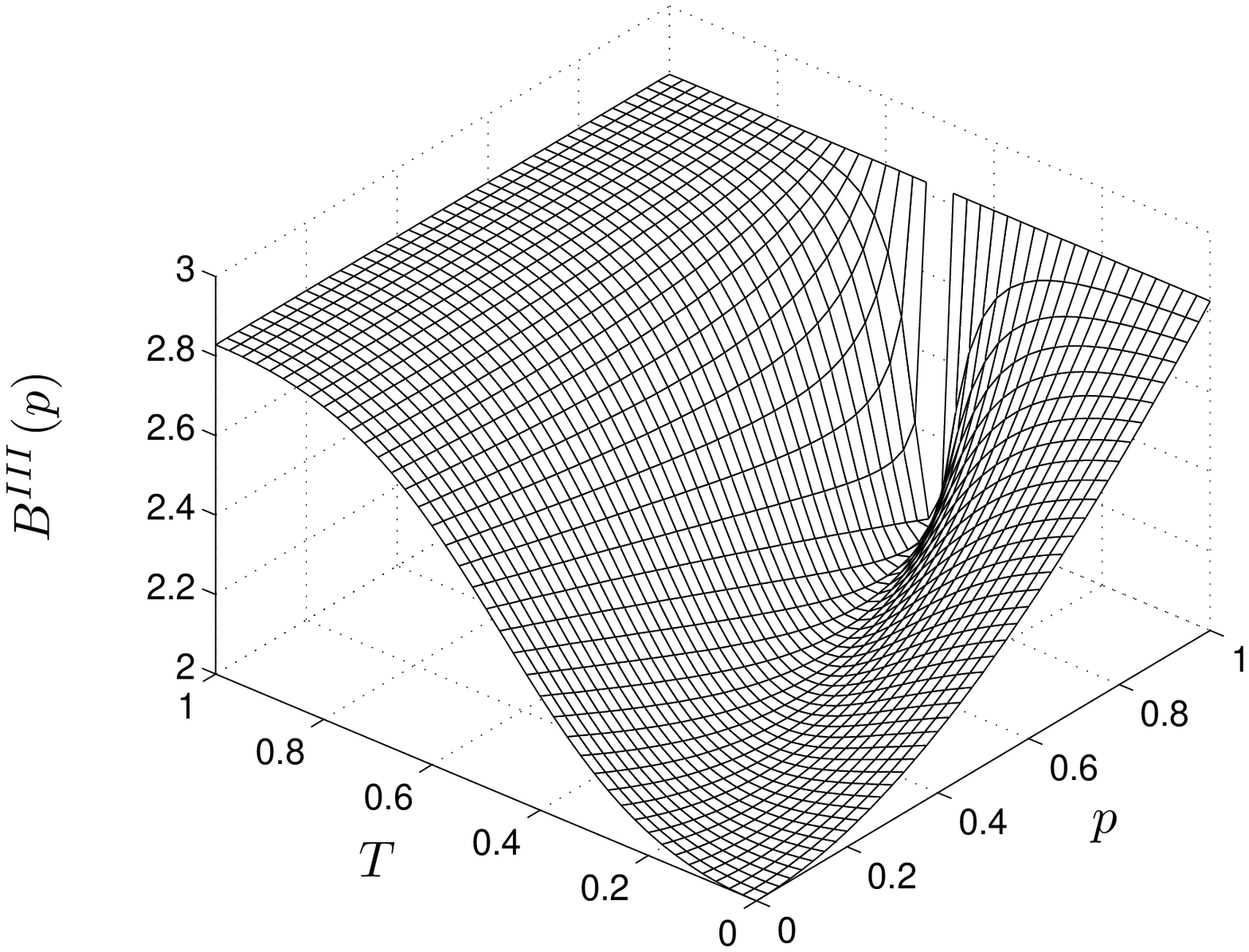}
\caption{Top plot shows maximal concurrence $C^{III}(p)$ and bottom plot
maximal violation of Bell inequalities $B^{III}(p)$ for the localized state.}
\label{concplusbell}
\end{figure}

For the mixture of both the distinguishable and indistinguishable
photons, the concurrence after the measurement is 
\begin{equation}
C^{I}(p)=\mbox{max}\left(0,\frac{|(1+p)T^2-pT|}{1-(2+p)T(1-T)}\right),
\end{equation} which is positive if $p\not= T/(1-T)$ and
\begin{equation} p<\frac{1-T}{T}+\frac{T}{1-T}.  
\end{equation}
Except $T=0,p/(1+p)$, the last condition is always satisfied
since right side has minimum equal to two. Even if the noise
is a mixture of distinguishable and indistinguishable photons,
it is possible to localize entanglement just by a measurement for
almost all the cases. After the application of LOCC polarization
filtering, the concurrence can be enhanced up to \begin{equation}
C^{III}(p,\epsilon)=\frac{2T^2|T-p(1-T)|}{\sqrt{1-2(1+p)(1-T)T}(\epsilon(1-T)^2+2T^2)}.
\end{equation} In the limit of
$\epsilon\rightarrow 0$, it approaches \begin{equation}
C^{III}(p)=\mbox{max}\left(0,\frac{|T-p(1-T)|}{\sqrt{1-2(1+p)T(1-T)}}\right)
\end{equation} which is plotted at the top of FIG.~\ref{concplusbell}. The maximal entanglement can be only induced for
$p=1$. Comparing both the discussed cases of the indistinguishable and distinguishable
photons, it is evident that it can be advantageous to induce
indistinguishability. For example, if they are distinguishable in time,
then spectral filtering can help us to make them more distinguishable and
consequently, the entanglement can be enhanced more by local filtering.

For partially indistinguishable photons given by parameter $p$,  we get the same value of the Bell factor 
after the measurement stage as after
the mixing stage $B^{II}(p)=B^{I}(p)$. 
In the range of parameters $p$ and $T$ where $B^{I}(p)<2$  
we are not able to violate
the Bell inequalities just by measurement although
the state just after the measurement changed its structure. 
The physical reason is that the difference between polarized
and unpolarized noise is not enough to guarantee violation of Bell
inequalities.
We can cross
the border of violation of Bell inequalities 
by performing local single-copy filtration operations generally on both
sides. 
We apply the two stage
filtration operations as stated previously. If the filtration parameter
\begin{widetext} \begin{equation}\label{cond}
  \epsilon \in\left(\frac{2 T^2 \left(\sqrt{1-2 (p+1) (1-T) T}-|
  T-p(1-T)|\right)}{(T-1)^2 \sqrt{1-2 (p+1) (1-T) T}},\frac{2 T^2
  \left(\sqrt{1-2 (p+1) (1-T) T}+| T-p(1-T)|\right)}{(T-1)^2 \sqrt{1-2
  (p+1) (1-T) T}} \right),
\end{equation} \end{widetext} then we get following Bell factor
\begin{equation}\label{eq_bell1}
  B^{III}(p,\epsilon)=\frac{4 \sqrt{2} T^2 |p (T-1)+T|}{\sqrt{2 (p+1) (T-1) T+1}
  \left(2 T^2+(T-1)^2 \epsilon\right)}
\end{equation} otherwise we get \begin{equation}\label{eq_bell2}
  B^{III}(p,\epsilon)=\frac{2 \sqrt{\frac{4 T^4 (p (T-1)+T)^2}{2 (p+1) (T-1)
T+1}+\left((T-1)^2 \epsilon-2 T^2\right)^2}}{2 T^2+(T-1)^2 \epsilon }.
\end{equation}
The particular cases of Bell factors given previously
can be obtained from the above relations setting $p=0$
($p=1$) for  coupling of distinguishable (indistinguishable) photons and additionally
the maximal value of the Bell factor is obtained if $\epsilon$ goes
to zero.  In order to get violation of the Bell inequalities
$B^{III}(p,\epsilon)>2$
the filtration parameter needs to be for the (\ref{eq_bell1}) $ \epsilon
< \frac{T^2 (p (T-1)+T)^2}{2 (T-1)^2 (2 (p+1) (T-1) T+1)}$ and for the
(\ref{eq_bell2}) $\epsilon < \frac{2 T^2 }{(T-1)^2}\left(\frac{\sqrt{2}
|p (T-1)+T|}{\sqrt{2 (p+1) (T-1) T+1}}-1\right)$. In both the cases,
value of $\epsilon$ should rapidly decrease as $T$ is smaller. In
other words, it means that for a given $\epsilon$, the state violates
Bell inequalities only if $p$ is sufficiently large. For any $p$
and $T$, there is always some $\epsilon>0$ below the region given by
(\ref{cond}), therefore as $\epsilon\rightarrow 0$, the maximal violation
\begin{equation}\label{belllimit}
  B^{III}(p)=2 \sqrt{1+\frac{(p (T-1)+T)^2}{2 (p+1) (T-1)
T+1}}.
\end{equation}
comes from the Eq.~(\ref{eq_bell1}) since it is higher
than the limit expression from Eq.~(\ref{eq_bell2}). The maximal violation
of Bell inequalities is plotted at the bottom of FIG.~\ref{concplusbell}. Generally, for $\epsilon>0$
we have to compare the (\ref{eq_bell1}) and (\ref{eq_bell2}) to find
out
which gives higher $B^{III}(p)$.


\section{Experimental Implementation}
Let us now describe the experimental implementation of the localization
protocol both for fully distinguishable and partially indistinguishable
photons.
\subsection{Experimental setup}
\begin{figure}
\centering
\includegraphics[scale=.32]{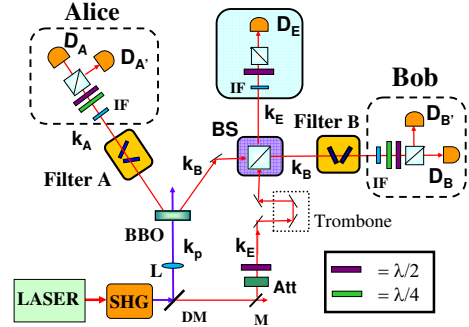}
\caption{(Color online) Schematic representation of the experimental setup. The dashed boxes indicate the polarization
analysis setup adopted by Alice and Bob.}
\label{FigSetup}
\end{figure}

The main source of the experiment is a Ti : Sa
mode-locked pulsed laser with wavelength (wl) $\lambda =795nm$, pulse
duration of $180fs$ and repetition rate $76MHz$: FIG.\ref{FigSetup}. A small portion of this
laser, breveted with a low reflectivity mirror $M$, generates the single
photon over the mode $k_{E}$ using an attenuator ($ATT$). The transformation
used to map the state $\left| H\right\rangle _{E}$ into $\rho _{E}=\frac{%
I_{E}}{2}$ is achieved either adopting a Pockels cell driven by a sinusoidal
signal, either through a stochastically rotated $\lambda /2$ waveplate
inserted on the mode $k_{E}$ during the experiment \cite{Scia04}. The main part of the
laser through a second harmonic generation (SHG) where a bismuth borate
(BiBO) crystal \cite{Ghot04}\ $\ $generates a UV laser beam having
wave-vector $k_{p}$ and wl $\lambda _{p}=397.5nm$ with power equal to $%
800mW$. A dichroic mirror (DM) separates the residual beam at $\lambda$ left after the SHG process from the UV laser beam. This field pump a $1.5mm$ thick non-linear crystal of $\beta $-barium borate (BBO) cut for type II phase-matching that generates
polarization maximally entangled pairs of photons. The
spatial and temporal walk-off is compensated by inserting a $\frac{\lambda }{%
2}$ waveplate and a $0.75$ mm thick BBO crystal on each output mode $k_{A}$
and $k_{B}$ \cite{Kwia95}. The photons of each pair are emitted with equal wavelength $%
\lambda =795nm$ over the two spatial modes $k_{A}$ and $k_{B}$.
In order to couple the noise with the signal, the photons on modes $k_{B}$
and $k_{E}$ are injected in the two input arms of an unbalanced
beam splitter $BS$ characterized by transmissivity $T$ and reflectivity
$R=1-T$. A mutual delay $\Delta t$, micrometrically adjustable by a
two-mirrors optical ''trombone'' with position settings $Z=2c\Delta t,$ can
change the temporal matching between the two photons. Indeed, the setting
value $Z=0$ corresponds to the full overlap of the photon
pulses injected into $BS$, i.e., to the maximum photon interference. On
output modes $k_{B^{\prime }}$ and $k_{E^{\prime }}$ the photons are
spectrally filtered adopting two interference filters (IF) with bandwidth
equal to $3nm$, while on mode $k_{A}$ the IF has a bandwidth of $4.5nm$.

Let us now describe how the measurement on the environment has been experimentally carried out.
According to the protocol described above, the photon propagating on mode $%
k_{E^{\prime }}$ is measured after a polarization analysis realized through
a $\frac{\lambda }{2}$ and a polarizing beam splitter (PBS).
In order to restore the optimum value of entanglement according to theory, a
filtering system has been inserted on $k_{A}$ and $k_{B^{\prime }}$ modes.
As shown in FIG.\ref{FigSetup}, the filtering is achieved by two sets of glass
positioned close to their Brewster's angle, in order to attenuate one
polarization in comparison to its orthogonal. The attenuation over the mode $%
k_{i}$ for the $H-$polarization ($V-$polarization) reads $A_{i}^{H}$ ($%
A_{i}^{V}$). By tuning the incidence angle, different values of attenuation
$\{A_{i}^{j}\}$ can be achieved.
Finally, to verify the working of the overall protocol, the emerging photons
on modes $k_{A}$ and $k_{B^{\prime }}$ are analyzed in polarization through
a $\frac{\lambda }{2}$ waveplate (wp), a $\frac{\lambda }{4}$ and a PBS.
Then the photons are coupled to a single mode fiber and detected by single
photon counting module (SPCM) $\{D_{A},D_{A'},D_{B'},D_{B}\}.$
The output signals of the detectors are sent to a coincidence box interfaced
with a computer, which collects the double coincidence rates
($[D_{A},D_{B}]$, $[D_{A},D_{B'}]$, $[D_{A'},D_{B}]$, $[D_{A'}
,D_{B'}]$) and triple coincidence rates
($[D_{A},D_{B},D_{E}]$, $[D_{A},D_{B'},D_{E}]$, $[D_{A'}
,D_{B},D_{E}]$, $[D_{A'},D_{B'},D_{E}]$). The detection of
triple coincidence ensures the presence of one photon per mode after the $%
BS$. In order to determine all the elements of the two-qubit density matrix, an overcomplete set of
observables is measured by adopting different polarization settings of the $%
\frac{\lambda }{2}$ wp and $\frac{\lambda }{4}$ wp positions \cite{Jame01}. The uncertainties on the different observables have
been calculated through numerical simulations of counting fluctuations due
to the Poissonian distributions. Here and after, we will
distinguish the experimental density matrix from the theoretical one by adding a "tilde" $\tilde{ }$.

As first experimental step, we have characterized the initial entangled
state generated by the NL crystal on mode $k_{A}$ and $k_{B}$, that is the signal to be transmitted through the noisy channel. The overall
coincidence rate is equal to about $8.000$ coincidences per second. The
experimental quantum state tomography of $\widetilde{\rho }_{AB}^{in}$ is
reported in FIG.\ref{ExpDist}-a, to be compared with the theoretical one $\rho
_{AB}^{in}=\left| \Psi^{in}\right\rangle _{AB}\left\langle \Psi^{in}\right|
_{AB}$, where $|\Psi^{in}\rangle=(|HV\rangle+i|VH\rangle)/\sqrt{2}$ (FIG.\ref{ThDist}-a). Although the generated state differs from the singlet state, all the conclusions from the previous section remains valid. We found a value of the concurrence $\tilde{C}_{in}=(0.869\pm
0.005)$ and a linear entropy $\tilde{S}_{in}=(0.175\pm 0.005)$, where
  the uncertainty on the concurrence has been calculated 
applying the Monte-Carlo method on the experimental density matrix. The fidelity with the
entangled state $\left| \Psi^{in}\right\rangle _{AB}$ is $\mathcal{F}(%
\widetilde{\rho }_{AB}^{in},\left| \Psi ^{in}\right\rangle
_{AB})=\left\langle \Psi ^{in}\right| _{AB}\widetilde{\rho }_{AB}^{in}\left|
\Psi ^{in}\right\rangle _{AB}=(0.915\pm 0.002)$. The discrepancy between
theory and experiment is mainly attributed to double pairs emission. Indeed
by subtracting the accidental coincidences we obtain a concurrence equal to $\widetilde{C}'_{in}=(0.939\pm0.006)$ and a linear entropy of $\widetilde{S}'_{in}=(0.091\pm0.006)$. The fidelity with the entangled state $\ket{Psi^{in}}_{AB}$ is then equal to $F'(\widetilde{\rho}^{in}_{AB},\ket{Psi^{in}}_{AB})=(0.949\pm0.003)$. Minor sources of imperfections related to the generation of our entangled state are the presence of spatial and temporal walk-off due to the non linearity of the BBO crystal as well as a correlation in wavelength of the generated pairs of photons.

\subsection{Coupling of distinguishable photons}
The experiment with distinguishable photons has been achieved by injecting a
single photon $E$ with a randomly chosen mutual delay with photon $B$ of $\Delta t>>\tau
_{ind}=300 fs.$ We note that the resolution time $t_{\det}$ of the detector is $t_{\det
}>>\Delta t,$ hence it is not technologically possible to individuate
whether the detected photon belongs to the environment or to the entangled
pair. To carry out the experiment we adopted a beam splitter with $T=0.40$ which,
according to the theoretical prediction (see FIG.~\ref{incoherentfig}),
allows to study the entanglement localization process.\\
Since multi-photon contributions represent the main source of imperfections in our calculations, we have subtracted their contributions in all the experimental data reported for the density matrices of each step of the protocol.

\textbf{I) Mixing}
 
\begin{figure}[h]
\centering
\includegraphics[scale=.43]{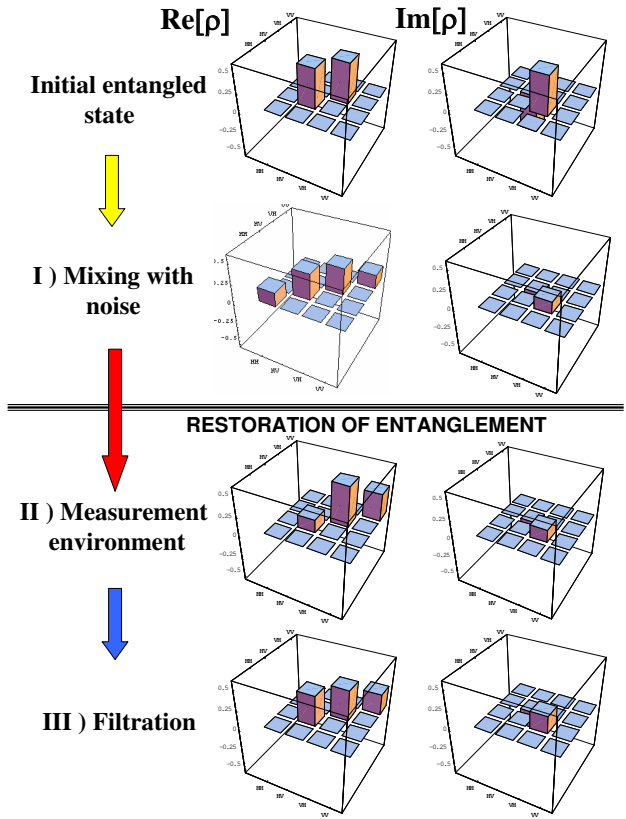}
\caption{(Color online) Theoretical density matrix for distinguishable photons,
enlightening the evolution of $\rho$
through each step of the entanglement localization protocol.}
\label{ThDist}
\end{figure}

\begin{figure}[h]
\centering
\includegraphics[scale=.43]{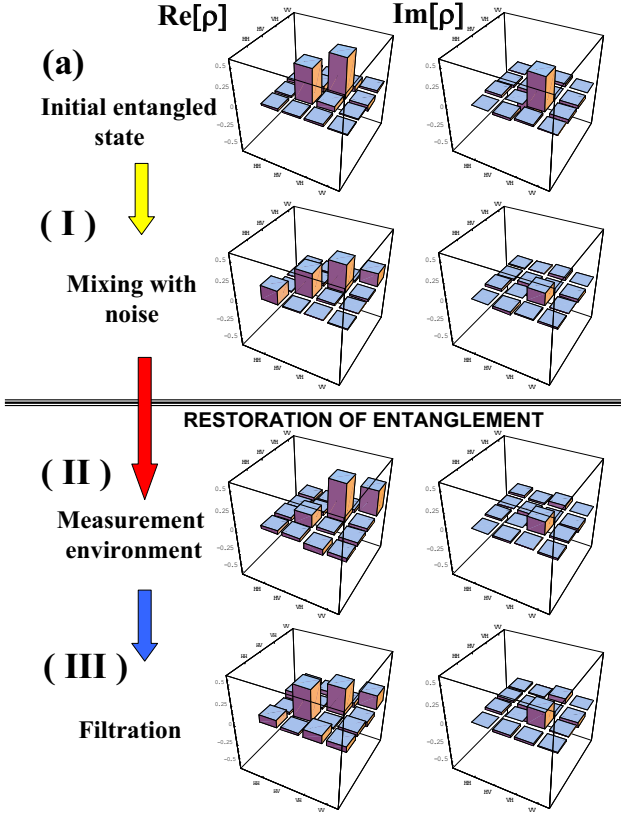}
\caption{(Color online) Experimental density matrix for distinguishable photons: \textbf{a}%
) $\widetilde{\protect\rho }^{in}$, \textbf{I}) $\widetilde{\protect%
\rho }^{I}_{dis}$, \textbf{II}) $\widetilde{\protect\rho }^{II}_{dis}$,
\textbf{III}) $\widetilde{\protect\rho }^{III}_{dis}. $
For each tomographic setting  the measurement lasts from 5 s ($\textbf{a}$) to 30 minutes ($\textbf{III}$), the last case corresponding to about $500$ triple coincidence. Contributions due to triple accidental coincidences have been subtracted from experimental data.}
\label{ExpDist}
\end{figure}

Without any access to the environmental photon, after the mixing on the $BS$, once there is one photon per output mode, the theoretical input state ${\rho^{in}}=\ket{\Psi^{in}}\bra{\Psi^{in}}$ evolves into a noisy state
represented by the density matrix ${\rho}^{I}_{dis}$, written in the basis $\{|HH\rangle ,|HV\rangle
,|VH\rangle ,|VV\rangle ,\}$ as:
\begin{equation}
  {\rho }^{I}_{dis}=\frac{1}{4P^{I}_{dis}}\left(
\begin{array}{cccc}
\alpha & 0 & 0 & 0 \\
0 & \beta & i\xi & 0 \\
0 & -i\xi & \gamma & 0 \\
0 & 0 & 0 & \delta
\end{array}
\right)
\end{equation}
 with $\alpha = \delta = R^{2}$, $\beta = \gamma =R^{2}+2T^{2}$,
 $\xi=2T^{2}$ and $P^{I}_{dis}=R^{2}+T^{2}$, where $R=1-T$ indicates
 the reflectivity of the beam splitter. This matrix, shown in
 FIG.(\ref{ThDist}-\textbf{I}), exhibits theoretically concurrence
 $C^{I}_{dis}=\max (0,\frac{2T^{2}-R^{2}}{2P_I})$. Because of the interaction with noise, $\rho^{I}_{dis}$ does not exhibit entanglement ($C_{dis}^I=0)$ and is highly mixed with linear entropy equal to $S_{dis}^I=0.90.$ Afterwards we analyze experimentally how the entangled state
 is corrupted due to the coupling with the photon $E$. In this case no-polarization selection
is performed on mode $k_{E}.$ The experimental density matrix $%
\widetilde{\rho}_{dis}^{I}$ is shown in FIG.(\ref{ExpDist}-\textbf{I}). As expected,
it exhibits $\widetilde{C}_{dis}^I=0,$ and $\widetilde{S}_{dis}^I=(0.89\pm 0.01)$. The fidelity with theory is high:
$\widetilde{F}_{dis}^{I}=F(\widetilde{\rho}_{dis}^{I},\rho^{I}_{dis})=(0.997\pm
0.006)$, with $F(\rho ,\sigma )=\text{Tr}^{2}\left[ \left( \sqrt{\rho }%
\sigma \sqrt{\rho }\right) ^{1/2}\right].$ In this case, due to the coupling the entanglement is completely redirected and no quantum distillation protocol can be applied to restore entanglement between photons $A$ and $B$.

\textbf{II) Measurement} 

As no selection has been performed on the
photon $E$ before the coupling, a multi-mode photon $E$
interacts with the entangled photon $B$. After the interaction, a single
mode fiber has been inserted on the output mode of the BS in order to
select only the output mode connected with the entanglement breaking.
Let us consider the case in which the photon on mode $k_{E}$
is measured in the state $|H\rangle_{E}$. 
After the measurement, the density matrix $\rho_{dis}^{I}$ evolves into an entangled one described by the density
matrix ${\rho}_{dis}^{II}$
 \begin{equation}
   {\rho }_{dis}^{II}=\frac{1}{4P^{II}_{dis}}\left(
\begin{array}{cccc}
0 & 0 & 0 & 0 \\
0 & \beta & i\xi & 0 \\
0 & -i\xi & \gamma & 0 \\
0 & 0 & 0 & \delta
\end{array}
\right)
\end{equation}
where $\delta=R^{2}$, $\beta=T^{2}$, $\gamma=R^{2}+T^{2}$,
$\xi=T^{2}$: FIG.(\ref{ThDist}-\textbf{II}). In this case the
concurrence reads $C^{II}_{dis}=\frac{T^{2}}{2P^{II}_{dis}}$ with
$P^{II}_{dis}=\frac{P^{I}_{dis}}{2}$. The entanglement is localized back for
all the values of $T\neq \{0,\frac{1}{2}\}$ but, as shown in the
graph, the elements of the matrix are fairly unbalanced. According to
theory, we expect a localization of the entanglement with
$C^{II}_{dis}=0.32$, and success rate $P^{II}_{dis}=0.27$.
Experimentally we obtain $\widetilde{C}^{II}_{dis}=(0.19\pm 0.02)>0$,
a fidelity with the theoretical density matrix ${\rho}^{II}_{dis}$
equal to $\widetilde{F}^{II}_{dis}=(0.96 \pm 0.03)$, while the mixedness of the
state decreases $\widetilde{S}^{II}_{dis}=(0.72\pm0.02)$ to be compared to theoretical prediction $S^{II}_{dis}=0.74$ . This is achieved at the cost of a probabilistic implementation
where the probability of success reads: $\widetilde{P}^{II}_{dis}=(0.26\pm 0.01)$. 

\textbf{III) Filtration} 

The filtration is introduced onto the mode $k_{A}$ and $k_{B}$, attenuating the vertical polarization, in order to increase the concurrence and symmetrizing the density matrix by
a lowering of the $|VV\rangle \langle VV|$ component in ${\rho}^{II}_{dis}$: FIG.(\ref{ThDist}-II). After filters $F_A$ and $F_B$, the density matrix reads:
\begin{equation}
  {\rho}^{III}_{dis}=
\left(
\begin{array}{cccc}
0 & 0 & 0 & 0 \\
0 & \epsilon\beta & i\epsilon\xi & 0 \\
0 & -i\epsilon\xi& \epsilon\beta & 0 \\
0 & 0 & 0 & \epsilon^{2}\delta
\end{array}
\right)
\end{equation}
where $\beta=T^{2}$, $\delta=R^{2},$
$\xi=\frac{T^{3}}{\sqrt{T^{2}+R^{2}}}$. The concurrence reads
$C^{III}_{dis}=\frac{2\epsilon
\xi}{2\epsilon\beta+\epsilon^{2}\delta}$ while the probability
of success of the protocol is equal to $P^{III}_{dis}=2\epsilon\beta+\epsilon^{2}\delta$. The intensity of filtration is quantified by the parameter $0<\epsilon\leq{1}$, connected to the attenuation of $V$ polarization. The concurrence
has a limit for asymptotic filtration ($\epsilon \rightarrow {0}$) lower
than unity and maximal entanglement cannot be approached. 
Of course, the collective protocols can be still used, since all entangled two-qubit state are distillable to a singlet one. By
setting the following attenuation values $A_{A}^{V}=0.33$,
$A_{A}^{H}=1$,
the theory predicts the achievement of $\rho^{III}_{dis}$ (FIG.(\ref{ThDist}-\textbf{III})) with the
mixedness $S^{III}_{dis}=0.76$ and the concurrence $C^{III}_{dis}=0.42$. The probability of success of
the filtration operation is $P^{III}_{dis}=0.43$ leading to an overall probability
of success of the entanglement restoration $P_{total}=P^{II}_{dis}P^{III}_{dis}=0.12.$ The
highest concurrence value which can be obtained with the coupling $T=0.40$
is equal to $C=0.55$ for $P_{total}\rightarrow 0$. Experimentally we
achieve the state shown in FIG.(\ref{ExpDist}-\textbf{III} with a fidelity
$\widetilde{F}^{III}_{dis}=(0.89 \pm 0.06)$ and measure a concurrence
equal to $\widetilde{C}^{III}_{dis}=(0.28\pm
0.02)>\widetilde{C}^{II}_{dis}$
while $\widetilde{P}^{III}_{dis}\widetilde{P}^{II}_{dis}=(0.11\pm 0.01)$ and $\widetilde{S}_{dis}^{III}=(0.70\pm0.02)$.
The experimental results shown above demonstrate the localization protocol validity, 
indeed a channel redirecting entanglement can be corrected to a channel
preserving relatively large amount of the entanglement.

Let's explain the choice of introducing experimentally just one set of filtration instead
of two. The glasses close to Brewster's angle attenuate the $V$
polarization. Let $\epsilon $ be the intensity of attenuation, the $%
|VH\rangle $ element of the density matrix will be attenuated by a factor $\sqrt{%
\epsilon }$. Even if the theory expects an
element $|HH\rangle $ close to zero, in the experimental case we obtain a
non-zero element due to the intensity of the coherent radiation, revealed by
double photons contributions. Hence a filtration of $\epsilon $ on both
modes, since it introduces an attenuation of $\epsilon $ on the element $%
|VV\rangle $ and of $\sqrt{\epsilon }$ on\ both the elements $|HV\rangle $
and $|VH\rangle $, would bring to a wider comparability between $|HH\rangle $
and $|HV\rangle $ elements, leading to a lower concurrence.

\subsection{Coupling of indistinguishable photons}

In order to exploit the regime where our localization protocol works best, we
let the signal photon and the noise one be spatially, spectrally and temporal
perfectly indistinguishable in the coupling process.
 A good spatial overlap is achieved by selecting the output modes through single mode fibers.

\textbf{I) Mixing.} 

\begin{figure}[h]
\centering
\includegraphics[scale=.43]{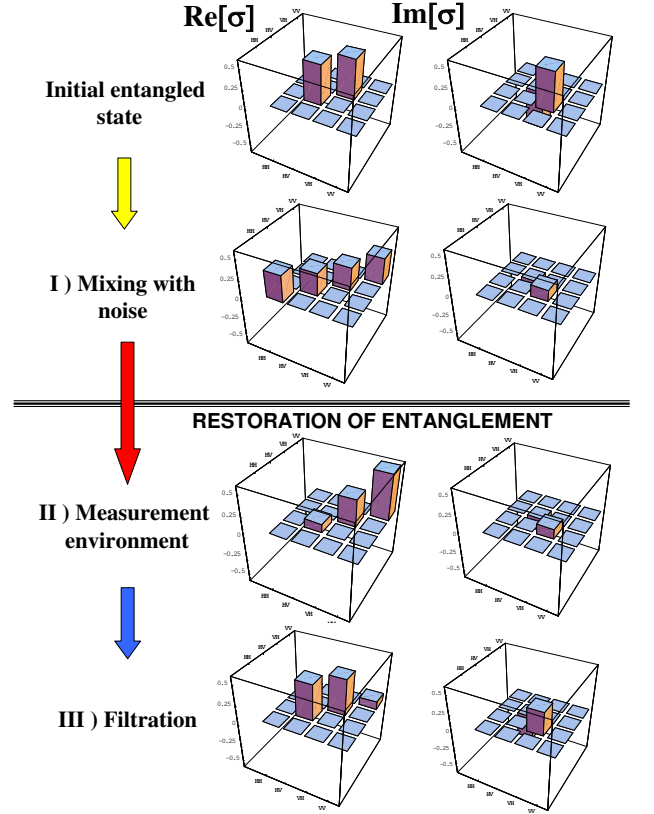}
\caption{(Color online) Theoretical density matrix for perfectly indistinguishable photons interacting on the $BS$.}
\label{ThInDist}
\end{figure}

In the fully indistinguishable photons regime, we indicate with
${\sigma}^{I}_{ind}$ the density matrix after the mixing on the $BS$, which is separable
for $T< 1/\sqrt{3}$, as shown in FIG.~\ref{coherentfig}.
The output state is a mixture of the initial state $|\Psi^{in}\rangle\langle\Psi^{in}|$ and the fully mixed one \cite{Wern89}: FIG.\ref{ThInDist}-I.
\begin{equation}
  \sigma^{I}_{ind}=\frac{1}{4P^{I}_{ind}}\left(
\begin{array}{cccc}
\alpha & 0 & 0 & 0 \\
0 & \beta & i\xi & 0 \\
0 & -i\xi & \gamma & 0 \\
0 & 0 & 0 & \delta
\end{array}
\right)
\end{equation}
with $\alpha = \delta = R^{2}$, $\beta = \gamma =T^{2}+(T-R)^{2}$,
$\xi=-T(T-R)$

\textbf{II) Measurement.} 

A measurement is carried out on the
environmental mode through the projection on $|H\rangle_{E^{\prime}}$. Thus the state
${\sigma}^{I}_{ind}$ evolves into ${\sigma}^{II}_{ind}$:
\begin{equation}
  \sigma^{II}_{ind}=\frac{1}{4P^{II}_{ind}}
\left(
\begin{array}{cccc}
0 & 0 & 0 & 0 \\
0 & \beta & i\xi & 0 \\
0 & -i\xi & \gamma & 0 \\
0 & 0 & 0 & \delta
\end{array}
\right)
\end{equation}
with $\beta=T^2$, $\gamma=T^2+(T-R)^2$, $\delta=R^2$, $\xi=-T(T-R)$.
The state is conditionally transformed into a maximally entangled
state (MEMS) \cite{Vers01b}: see FIG.(\ref{ExpInDist}-II). In
particular, it is found $C^{II}_{ind}=\frac{T|T-R|}{2P^{II}_{ind}}$
and $P^{II}_{ind}=\frac{T^2+(T-R)^2+R^2}{4}$.

\textbf{III) Filtration} 

Due to the strong asymmetry of
${\sigma}^{II}_{ind}$ a filter is introduced either on Alice or Bob mode, depending on the value of $T$. If $T>|2T-1|$, filter $F_B$ acts on Bob mode, as $\ket{H}_{B}\rightarrow{\frac{2T-1}{T}\ket{H}_{B}}$, while if $T<|2T-1|$, the filter acts on Alice mode, as $\ket{V}_{A}\rightarrow{\frac{T}{2T-1}\ket{V}_{A}}$.
In order to increase the concurrence, a second filter is inserted on
both Alice and Bob mode, which attenuates the vertical polarization
component: $\ket{V}\rightarrow \sqrt{\epsilon}\ket{V}$. The
filters $F_A$ and $F_B$ transform the density matrix into
$\sigma^{III}_{ind}$:
\begin{equation}
  {\sigma}^{III}_{ind}=\frac{1}{4P^{III}_{ind}}
\left(
\begin{array}{cccc}
0 & 0 & 0 & 0 \\
0 & \epsilon\alpha & i\epsilon\alpha & 0 \\
0 & -i\epsilon\alpha & \epsilon\alpha & 0 \\
0 & 0 & 0 & \epsilon^{2}\delta
\end{array}
\right)
\end{equation}
where $\alpha=T^{2}$, $\delta=\left( \frac{TR}{R-T}\right) ^{2},$ for
$T<|2T-1|$ and $\alpha=(2T-1)^{2}$, $\delta=R ^{2},$ for $T>|2T-1|$:
FIG.\ref{ThInDist}-II. The concurrence has value
$C^{III}_{ind}=\frac{2\epsilon\alpha}{2\epsilon\alpha+\epsilon^{2}\delta}$
while the probability reads $P^{III}_{ind}=\frac{2\epsilon\alpha+\epsilon^{2}\delta}{4}$.
Hence in the limit of asymptotic filtration ($\epsilon \rightarrow 0$), the concurrence reaches unity except for $T=1/2$ \cite{Vers01b}.

\subsection{Coupling of partially indistinguishable photons}

Let us now face up to a model which contemplates a situation close to the
experimental one. In fact a perfect indistinguishability between the photons $B$ and $E$ is almost impossible to achieve experimentally. We consider the density matrix ${\tau}_{AB}$ of the state shared between Alice and
Bob after the coupling, as a mixture arising from coupling with a partially distinguishable noise photon. The degree of indistinguishability is parametrized by the probability $p$ that the fully depolarized photon $E$ is completely indistinguishable from the photon $E$.

In order to maximize the degree of indistinguishability between the photon of modes $%
k_{B}$ and $k_{E}$ mixed on the beam splitter $BS$ we adopt narrower
interference filters and single mode fibers on the output modes. Hence we insert
interference filters on mode $k_{B\prime}$ and $k_{E\prime}$, with $\Delta\lambda=1.5nm$. Referring to the
previous nomenclature, the density matrix $\tau$ can be written as:
${\tau}_{AB}=p{\sigma}_{ind}+(1-p) {\rho}_{dis}$.
As first step, we have estimated the degree of indistinguishability between
the photon belonging to mode $k_{B}$ and the one associated to $k_{E}$. This
measurement has been carried out by realizing an Hong-Ou-Mandel
interferometer \cite{Hong87} adopting a balanced $50:50$ beam splitter
instead of $BS$. The photons on modes $k_{B^{\prime }}$ and $%
k_{E^{\prime }}$ are measured in the same polarization state $\left|
H\right\rangle $. The visibility of the Hong-Ou-Mandel dip has been measured
to be $V=(0.67 \pm  0.02)$. By subtracting estimated double pairs
contributions to the three-photon coincidence we have obtained a value of the
visibility $(0.75\pm 0.02)$. We attribute the mismatch with the
unit value to a different spectral profile between the coherent beam and the
fluorescence, which induces a distinguishability between photons on the input
modes of the beam splitter. From the visibility of the dip, we estimate a value of
$p$ equal to $(0.85\pm 0.05)$. Hence we have checked the Hong-Ou-Mandel interference with the unbalanced
beam splitter $BS$, and we have obtained a visibility of $(0.40\pm0.02)$. This result has been compared to a theoretical visibility $V$
previously calculated considering the expression $V=p\frac{2RT}{R^{2}+T^{2}}%
=(0.62\pm0.05)$ with $p$ as indicated by the Hong-Ou-Mandel
experiment, and $T=0.3$. To carry out the experiment the optical delay has been set in the position $\Delta t=0$. The mismatch between the two visibilities are due to multi-photon contributions. Indeed by taking into account the accidental coincidence we have estimated $V=(0.49\pm0.03)$.

\textbf{I) Mixing.} 

Analogously to what has been showed in the distinguishable case, after the mixing on the $BS$, 
once there is one photon per output mode, the input state evolves into a noisy state 
represented by the density matrix $\widetilde{\tau }^{I}$. 
The experimental density matrix is characterized by a fidelity with theory : $%
F(\widetilde{\tau }^{I},\tau^{I} )=(0.86\pm 0.02)$ and vanishing concurrence ($\widetilde{C}_I=0$).

\textbf{II) Measurement.} 

\begin{figure}[t]
\centering
\includegraphics[scale=.35]{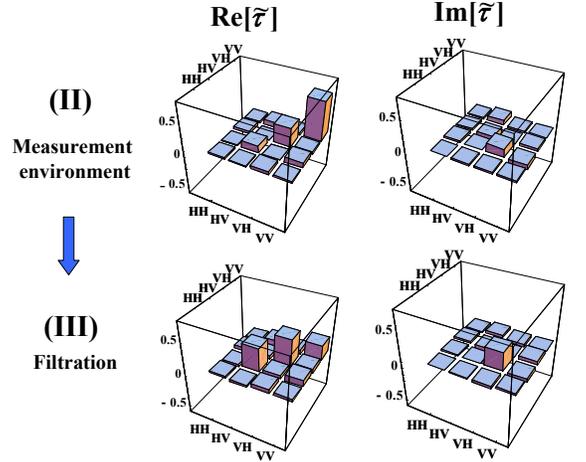}
\caption{(Color online) Experimental density matrix for partially indistinguishable photons:$\widetilde{\tau}$ with $p=0.85$. \textbf{II}) $\widetilde{\tau }^{II}$,
\textbf{III}) $\widetilde{\tau}^{III}. $ }
\label{ExpInDist}
\end{figure}

After measuring the photon on mode $k_{E^{\prime }}$, the
density matrix $\widetilde{\tau }^{I}$ evolves into $\widetilde{%
\tau }^{II}$: FIG.(\ref{ExpInDist}-\textbf{II}). The entanglement is localized with a
concurrence equal to $\widetilde{C}_{II}=(0.15\pm 0.03)>0$ to be compared with $C_{II}=0.22$; in this case the probability of success reads $\widetilde{P}_{II}=(0.22\pm 0.01)$, while theoretically we expect $P_{II}=0.2$. The fidelity with the theoretical state is $F(\widetilde{\tau }^{II},\tau
^{II})=(0.96\pm 0.01).$ 

\textbf{III) Filtration.}

Applying experimentally the filtration with the parameters $A_{A}=0.12$, and $A_{B}=0.30$
we obtain the state shown in FIG.(\ref{ExpInDist}-\textbf{III}). Hence we measure a higher
concurrence $\widetilde{C}_{III}=(0.50\pm 0.10)>\widetilde{C}_{II}$ while the expected theoretical value is $C_{III}=0.47$. The filtered state has $F(\widetilde{\tau
}^{III},\tau^{III})=(0.92\pm 0.04),$ and is post-selected with an overall success rate equal to $P_{total}=\widetilde{P}_{III}\widetilde{P}_{II}=(0.10\pm 0.01)$, where theoretically $P_{III}P_{II}=0.09$. This is a clear experimental demonstration of how an induced indistinguishability enhances the localized concurrence.

\section{Conclusion}

In summary, we have discussed the role of distinguishability in an elementary
polarization entanglement three-photon localization protocol, both 
theoretically and experimentally, especially for the cases of
total loss of entanglement. Theoretically, a full indistinguishability of
noisy surrounding photon and one photon from the maximally entangled
state of two photons is a necessary condition for perfect entanglement
localization. Moreover, also for the partially indistinguishable or even fully
distinguishable photon the localization protocol is still able to restore
non-zero entanglement. Using local single-copy polarization filters,
the entanglement can be enhanced even to violate the Bell
inequalities. The generalization of the results to linear polarization
sensitive coupling is also enclosed. Experimentally, the localization
of entanglement is demonstrated for the coupling between fully
distinguishable photons as well
as for the highly (but still partially) indistinguishable photons. The restoration of
entanglement after its previous break
was experimentally verified, also the positive role of the
indistinguishability induced by the spectral filters has been experimentally
checked. These results clearly show the broad applicability of the basic
element of the polarization-entanglement localization protocol in
realistic situations, where the surrounding uncontrollable noisy
system exhibiting just moderate indistinguishability of photons completely destroys the
entanglement due to the coupling process. In these cases, the collective
entanglement distillation method without a measurement on the
surrounding system can not restore any entanglement out of the
coupling. 

In this paper we extensively discuss the simplest proof-of-principle
entanglement localization method with a novel focus just to the effect
caused by the indistinguishability of the single surrounding photon. The
entanglement
localization after the multiple single interactions with the photons
is a natural step of further investigation. An open question is whether
after such multiple decoherence events on both sides of the entangled
state, non-zero entanglement can be as well always localized back just by
local measurements and classical communication, or a collective
localization measurement will be required. Further, an extension from
fixed number of surrounding particles to randomly fluctuating number
of the surrounding particles will be interesting. Another interesting
direction is to analyze the entanglement localization after another kind of the
coupling of partially indistinguishable objects, especially between atoms/ions and light
or
between individual atoms or ions. This gradual investigation will give
a
final answer to an important physical question: how to manipulate
quantum
entanglement distributed by the coupling to noisy (in)distinguishable complex
environments.


\section{Appendix A}

To describe a linear coupling (beam splitter) with different transmissivity
for vertical and horizontal polarization, we use the amplitude
transmissivities $t_v$ and $t_h$. The intensity transmissivities are then
$T_V=t_v^2$ and $T_H=t_h^2$ and the previous result can be obtained
taking $T=T_V=T_H$.

\subsection{Coupling of indistinguishable photons}

After the mixing of indistinguishable photons $B$ and $E$
on the beam splitter, without any access to the photon $E$, the single
state transforms to the mixed state (if two photons leave separately)
exhibiting the concurrence
\begin{equation}
  C_{ind}^{I}=\max\left(0,\frac{2t_{v}t_{h}|t_{h}^{2}-1+t_{v}^2|-(1-t_{v}^{2})
  (1-t_{h}^{2})}{2P_{ind}^{I}}\right),
\end{equation}
where
$P_{ind}^{I}=(2t_{v}^2t_{h}^2+(1-2t_{v}^2)^2+(1-2t_{h}^2)^2+2(1-t_{v}^2)(1-t_{h}^2))/4$
is the success probability. The concurrence is depicted on FIG.~9, there
is evidently a large area where the concurrence vanishes completely and
the entanglement is lost between $A$ and $B$.

\begin{figure} \centerline{\psfig{height=8cm,angle=270,file=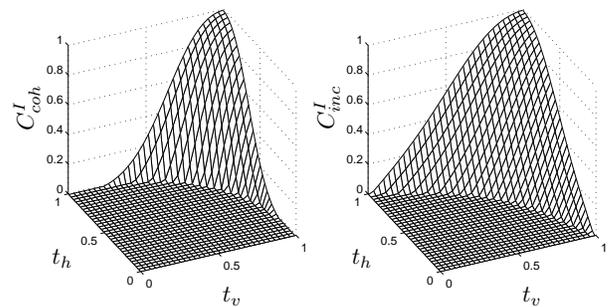}}
\caption{Concurrence before the measurement for indistinguishable
(left) and
distinguishable (right) couplings.}
\end{figure}

\begin{figure} 
\centerline{\psfig{height=8cm,angle=270,file=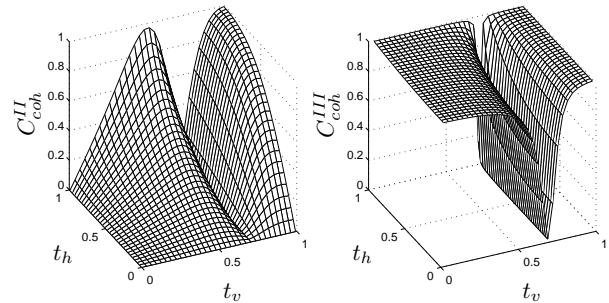}}
\caption{Concurrence after the measurement (left) and after the filtration
with $\epsilon=0.1$ (right) for indistinguishable photons
(indistinguishable coupling).}
\end{figure}
To restore the entanglement a general projection measurement can be
considered on the  photon $E$. But it is too complex to find an optimal
projection analytically, rather the optimal measurement can be find
numerically, for the particular values of $t_v$ and $t_h$. On the other
hand, it is possible to find a sufficient condition to restore the
entanglement for some particular orientation of the measurement.  Let us
assume the projection of the photon $E$ on the state $|V\rangle$. Then
the state $\sigma_{ind}$ changes
and exhibits the concurrence
\begin{equation}
  C_{ind}^{II}=\frac{t_{h}t_{v}|1-2t_{v}^{2}|}{2P_{ind}^{II}},
\end{equation}
where $P_{ind}^{II}=(t_{v}^2t_{h}^2+(1-2t_{v}^2)^2+(1-t_{v}^2)(1-t_{h}^2))/4$
is the success rate. Such state is always entangled
except $t_h=0$ or $t_v^2=1/2,0$ and can be
further filter out by the filter F1 on photon B which makes
$|V\rangle\rightarrow\frac{2t_{v}^2-1}{t_{v}t_{h}}|V\rangle$ for
$t_h>\frac{|2t_{v}^2-1|}{t_{v}}$ or by the filter F1 on photon
A $|H\rangle\rightarrow\frac{t_{v}t_{h}}{2t_{v}^2-1}|H\rangle$
for $t_h<\frac{|2t_{v}^2-1|}{t_{v}}$ and subsequently,
the filter F2 attenuating horizontal polarization
$|H\rangle\rightarrow\sqrt{\epsilon}|H\rangle$, applied to both 
photons A and B. As result, the state is transformed by both types
of filtering to a new state
with the corresponding concurrence
\begin{equation}
  C^{III}_{ind}(\epsilon)=\frac{\epsilon(1-2t_{v}^{2})^2}{2P_{coh_{1}}^{III}}=\frac{\epsilon
  t_{v}^{2}t_{h}^{2}}{2P_{coh_{2}}^{III}},
\end{equation}
where $P_{coh_{1}}^{III}=(2\epsilon
(1-2t_{v}^2)^2+\epsilon^2(1-t_{v}^2)(1-t_{h}^2))/4$ is the success rate of
the localization using the first type of filtration F1 on photon B and
$P_{coh_{2}}^{III}=(2\epsilon
(t_{v}t_{h})^2+\epsilon^2(1-t_{v}^2)(1-t_{h}^2)/(1-2t_{v}^{2})^{2})/4$
is the success rate of the restoration using the second type of filtration on photon
A. The concurrence approaches unity as $\epsilon\rightarrow
0$.
The similar results can be obtained for the projection of the photon $E$
on the state $|H\rangle$, only the substitution $t_h\leftrightarrow t_v$
is necessary. In conclusion, except the cases $t^2_v,t^2_h=1/2$, the
entanglement can be always localized and enhanced arbitrarily close to
maximal pure entangled state, at cost of the success rate, similarly
to previous analysis. The concurrences after the measurement and the one after the filtration in the localization procedure are
depicted on FIG.~10.

\subsection{Coupling of distinguishable photons}

For distinguishable photons, without any access to
the photon $E$, after the mixing at the beam splitter the singlet state
transforms to a state
exhibiting the concurrence
\begin{equation}
  C^{I}_{dis}=\max\left(0,\frac{t_{h}t_{v}
  (t_{h}^{2}+t_{v}^{2})-(1-t_{v}^{2})(1-t_{h}^{2})}{2P_{dis}^{I}}\right),
\end{equation}
where $P_{dis}^{I}=(2(1-t_{v}^{2})(1-t_{h}^{2})+(1-t_{h}^{2})^2+
t_{h}^{2}(t_{h}^{2}+t_{v}^2)+(1-t_{v}^{2})^2+t_{v}^{2}(t_{h}^{2}+t_{v}^2))/4$
is the probability of success. Similarly, there is a large area of the
parameters in which the concurrence vanishes and the entanglement is
lost between $A$ and $B$, as can be seen from FIG.~9.

 \begin{figure}
{\psfig{height=8cm,angle=270,file=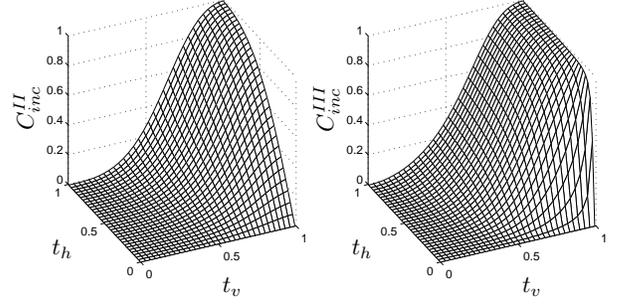}} 
\caption{Concurrence after
the measurement (left) and after the filtration with $\epsilon=0.1$
(right) for distinguishable coupling.}
 \end{figure}

 \begin{figure}
{\psfig{height=8cm,angle=270,file=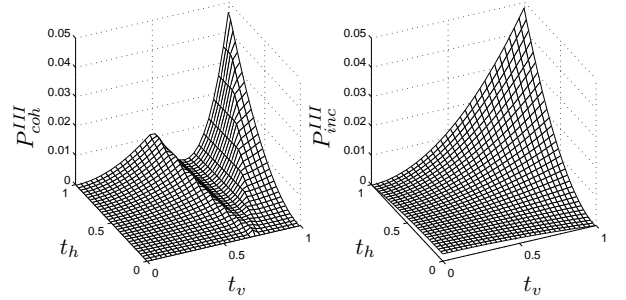}} 
\caption{Success rate after
the filtration with $\epsilon=0.1$ for indistinguishable photons (left)
and distinguishable photons (right)} \end{figure}

Similarly, it is sufficient to localize entanglement by the projection of
the environmental photon on the state $|V\rangle$. After the successful
projection, the mixed state
has concurrence
\begin{equation}
  C_{dis}^{II}=\frac{t_{h}t_{v}^3}{2P_{dis}^{II}},
\end{equation}
where
$P_{dis}^{II}=(t_{h}^2t_{v}^2+t_{v}^4+(1-t_{v}^2)^2+(1-t_{v}^{2})(1-t_{h}^{2}))/4$
is the success rate of the projection. Except trivial
$t_v,t_h=0$, the entanglement is always localized by the
projection on the environment. But for this case, optimal
filtration by F1 and F2 defined as $|V\rangle\rightarrow
\frac{t_{v}t_{h}}{\sqrt{t_{v}^4+(1-t_{v}^2)^2}}|V\rangle$ and
$|H\rangle\rightarrow\sqrt{\epsilon}|H\rangle$ produces the state
with overall probability of success $P_{dis}^{III}=(2\epsilon
t_{v}^2t_{h}^2+\epsilon^2(1-t_{v}^{2})(1-t_{h}^{2}))/4$, which gives
the concurrence
\begin{equation}
  C_{dis}^{III}(\epsilon)=\frac{\epsilon t_{v}^4 t_{h}^2}{2P_{dis}^{III}\sqrt{t_{v}^4+(1-t_{v}^2)^2}}
\end{equation}
In the limit $\epsilon\rightarrow 0$ it gives the maximal concurrence
\begin{equation}
  C_{dis}^{III}=\frac{t_{v}^2}{\sqrt{t_{v}^4+(1-t_{v}^2)^2}}.
\end{equation}
In a similar way as in the symmetrical case, the maximal entanglement cannot
be approached if the photon $E$ is in principle distinguishable. The
concurrence before and after the filtration is depicted on FIG.~11.
Success rates after the filtration for both the indistinguishable and
distinguishable photons are plotted in FIG.~12.

\medskip
\noindent {\bf Acknowledgments}

The research has been supported by
projects  No. MSM 6198959213 and No. LC06007 of the Czech Ministry
of Education. R.F. acknowledges grant 202/08/0224 of GA CR and support by the Alexander von Humboldt
Foundation.



\begin{thebibliography}{99}



\bibitem{Benn96} C. H. Bennett, G. Brassard, S. Popescu, B. Schumacher,
J. A. Smolin, and W. K. Wootters, \textit{Phys. Rev. Lett.} \textbf{76}, 722 (1996).

\bibitem{Deut96}D. Deutsch, A. Ekert, R. Jozsa, Ch. Macchiavello, S. Popescu and Anna
Sanpera, \textit{Phys. Rev. Lett.} \textbf{77}, 2818 (1996). 

\bibitem{Zhao03} Z. Zhao, T. Yang, Y. A. Chen, A. N. Zhang, and J. W. Pan, \textit{Phys. Rev. Lett.} \textbf{90} 207901 (2003).

\bibitem{Pan03} J. W. Pan, S. Gasparoni, R. Ursin, G. Weihs and A. Zeilinger, Nature
\textbf{423}, 417 (2003).

\bibitem{Walt05} P. Walther, K. J. Resch, È. Brukner, A. M. Steinberg,
J. W. Pan, and A. Zeilinger, \textit{Phys. Rev. Lett.} \textbf{94} 040504 (2005).

\bibitem{Kwia01} P. G. Kwiat, S. Barraza-Lopez, A. Stefanov, and N. Gisin, Nature (London) \textbf{409}, 1014 (2001).

\bibitem{Cohe98} O. Cohen, \textit{Phys. Rev. Lett.} \textbf{80}, 2493 (1998).

\bibitem{DiVi04} D. P. DiVincenzo et al, Lecture Notes in Computer Science 1509 (Springer-
Verlag, Berlin, 1999), pp. 247-257 (2004).

\bibitem{Vers04} F. Verstraete, M. Popp and J.I. Cirac, \textit{Phys. Rev. Lett.}
\textbf{92}, 027901 (2004).

\bibitem{Smol05} J. A. Smolin, F. Verstraete, and A. Winter, \textit{Phys. Rev.
A} \textbf{72}, 052317 (2005).

\bibitem{Popp06} M. Popp, F. Verstraete, M. A. Martin-
Delgado and J. I. Cirac, \textit{Phys. Rev. A} \textbf{71}, 042306 (2005).

\bibitem{Gour06} G. Gour and R.W. Spekkens, \textit{Phys. Rev. A} \textbf{73}, 062331 (2006).

\bibitem{Zure03} W. H. Zurek, \textit{Rev. Mod. Phys.} \textbf{75}, 715 (2003).

\bibitem{Sciar09} F. Sciarrino, E. Nagali, F. De Martini, M. Gavenda, and R. Filip, \textit{Phys. Rev. A} \textbf{79}, 060304(R) (2009).

\bibitem{Horo97} M. Horodecki, P. Horodecki and R. Horodecki, \textit{Phys. Rev. Lett.} \textbf{78}, 574 (1997).

\bibitem{Lind98} N. Linden, S. Massar, and S. Popescu, \textit{Phys. Rev. Lett.} \textbf{81}, 3279 (1998). 

\bibitem{Kent99} A. Kent, N. Linden, and S. Massar, \textit{Phys. Rev. Lett.} \textbf{83}, 2656 (1999). 

\bibitem{Horo99} M. Horodecki, P. Horodecki, and R. Horodecki, \textit{Phys. Rev. A} \textbf{60}, 1888 (1999). 

\bibitem{Vers01} F. Verstraete, J. Dehaene and B. DeMoor, \textit{Phys. Rev. A} \textbf{64},
010101 (2001).
 
\bibitem{Cen02} L. X. Cen, N. J. Wu, F. H. Yang, and J. H. An, \textit{Phys. Rev. A} \textbf{65} 052318 (2002).
 
\bibitem{Vers02} F. Verstraete and M. M. Wolf, \textit{Phys. Rev. Lett.} \textbf{89} 170401 (2002).
 
\bibitem{Pete04} N. A. Peters, J. B. Altepeter, D. Branning, E. R. Jeffrey, T. C. Wei, and P. G. Kwiat, \textit{Phys. Rev. Lett.} \textbf{92}, 133601 (2004). 

\bibitem{Wang06} Z. W. Wang, X. F. Zhou, Y. F. Huang, Y. S. Zhang, X. F. Ren, and G. C. Guo, \textit{Phys. Rev. Lett.} \textbf{96} 220505 (2006). 

\bibitem{Lian08} Y. C. Liang, L. Masanes, and A. C. Doherty, Phys. Rev. A 77 012332 (2008).

\bibitem{Bell64} J. S. Bell, Physics 1, 195 (1964).

\bibitem{Clau69} J. F. Clauser, M. A. Horne, A. Shimony and R. A. Holt, \textit{Phys. Rev. Lett.} \textbf{23}, 80 (1969).




\bibitem{Wern89} R.F. Werner, \textit{Phys. Rev. A} \textbf{40}, 4277 (1989).

\bibitem{Woot98} W. Wootters, \textit{Phys. Rev. Lett.} \textbf{80}, 2245 (1998).

\bibitem{Horo95} R. Horodecki, P. Horodecki and M. Horodecki, \textit{Phys. Lett. A} \textbf{200}, 340 (1995).





\bibitem{Ishi00} S. Ishizaka and T. Hiroshima, \textit{Phys. Rev. A} \textbf{62}, 022310 (2000).

\bibitem{Wei03}  T. C. Wei, K. Nemoto, P. M. Goldbart, P. G.
Kwiat, W. J. Munro, and F. Verstraete, \textit{Phys. Rev. A} \textbf{67}, 022110
(2003).

\bibitem{Cine04} C. Cinelli, G. Di Nepi, F. De Martini, M. Barbieri, and P.
Mataloni, \textit{Phys. Rev. A} \textbf{70}, 022321 (2004).

\bibitem{Scia04} F. Sciarrino, C. Sias, M. Ricci, and F. De Martini, \textit{Phys. Rev. A} \textbf{70}, 052305 (2004).

\bibitem{Ghot04} M. Ghotbi, M. Ebrahim-Zadeh, A. Majchrowski, E. Michalski,
I.V. Kityk, \textit{Optics Letters} \textbf{29}, 21 (2004).

\bibitem{Kwia95} P. G. Kwiat, K. Mattle, H. Weinfurter, A. and Zeilinger, \textit{Phys. Rev. Lett.} \textbf{75}, 4337 (1995).

\bibitem{Jame01} D. F. V. James, P.G. Kwiat, W. J. Munro, and A. G. White, \textit{Phys. Rev. A} \textbf{64}, 052312 (2001).

\bibitem{Vers01b} F. Verstraete, K. Audenaert,  T. De Bie, and B. De Moor, \textit{Phys. Rev A} \textbf{64}, 012316 (2001).

\bibitem{Hong87} C K.Z. Hong, Y. Ou, and L. Mandel, \textit{Phys. Rev. Lett.} \textbf{59}, 2044 (1987).

\end{thebibliography}
\end{document}